\documentclass[preprint,12pt]{aastex}
\def\gs{$\Gamma_{\rm soft}$}
\def\civ{{\sc{Civ}}$\lambda$1549\/}
\def\civbc{{\sc{Civ}$_{BC}$}$\lambda$1549\/}
\def\ltsima{$\; \buildrel < \over \sim \;$}
\def\simlt{\lower.5ex\hbox{\ltsima}}
\def\hb{{\sc{H}}$\beta$\/}
\def\hbbc{{\sc{H}}$\beta_{\rm BC}$\/}

\received{} \accepted{}
\def\ne{n$_{\rm e}$\/}

\def\rfe{R$_{\rm FeII}$}
\def\feiiq{\rm Fe{\sc ii }$\lambda$4570\/}

\def\ltsima{$\; \buildrel < \over \sim \;$}
\def\simlt{\lower.5ex\hbox{\ltsima}}            % < over MMM
\def\gtsima{$\; \buildrel > \over \sim \;$}

\def\simgt{\lower.5ex\hbox{\gtsima}}            % > over MMM

\def\ha{{\sc H}$\alpha$}
\def\civ{{\sc{Civ}}$\lambda$1549\/}

\def\civbc{{\sc{Civ}}$\lambda$1549$_{\rm BC}$\/}

\def\cm3{cm$^{-3}$\/}
\def\hb{{\sc{H}}$\beta$\/}

\def\hbbc{{\sc{H}}$\beta_{\rm BC}$\/}

\def\ciii{{\sc{Ciii]}}$\lambda$1909\/}
\def\oiiiopt{{\sc{[Oiii]}}$\lambda$4959,5007\/}
\def\o4363{{\sc{[Oiii]}}$\lambda$4363\/}

\def\siiii{\ion{Si}{3}]$\lambda$1892\/}

\def\feiiopt{{Fe \sc{ii}}$_{\rm opt}$\/}
\def\feii{{Fe\sc{ii}}$_{\rm opt}$\/}
\def\fe{{\sc{Fe}}\/}

\def\fe76087{{\sc [Fe vii]}$\lambda$6087\/}
\def\oiii{{\sc [Oiii]}$\lambda$5007}

\def\kms{km~s$^{-1}$}

\begin{document}
\title{\sc Searching for the Physical Drivers of the Eigenvector 1 Correlation
Space}

\author{P. Marziani\altaffilmark{1}, J. W. Sulentic\altaffilmark{2},   T.
Zwitter\altaffilmark{3},
 D. Dultzin-Hacyan\altaffilmark{4}, and M. Calvani\altaffilmark{1}}

\altaffiltext{1}{Osservatorio Astronomico di Padova, vicolo
dell'Osservatorio 5, I$-$35122 Padova, Italy;
marziani@pd.astro.it, calvani@pd.astr.it }

\altaffiltext{2}{Department of  Physics and Astronomy, University
of Alabama, Tuscaloosa, AL 35487; giacomo@merlot.astr.ua.edu}

\altaffiltext{3}{Department of Physics, University of Ljubljana,
Jadranska 19, 1000, Ljubljana, Slovenia; tomaz.zwitter@uni-lj.si}

\altaffiltext{4}{Instituto de Astronom\'\i a, UNAM, Mexico, DF
04510, Mexico; deborah@astroscu.unam.mx}
\begin{abstract}

We recently discussed an Eigenvector 1 (E1) parameter space that
provides optimal discrimination between the principal classes of
broad-line Active Galactic Nuclei (AGN).  In this paper we begin a
search for the most important physical parameters that are likely
to govern correlations and data point distribution in E1 space. We
focus on the principal optical parameter plane involving the
width of the \hb\ broad component  [FWHM(\hbbc)] and the
equivalent width ratio between the  Fe {\sc ii} blend at
$\lambda$4570 and \hbbc. We show that the observed correlation
for radio-quiet sources can be accounted for if it is primarily
driven by the ratio of AGN luminosity to black hole mass (L/M
$\propto$ Eddington ratio) convolved with source orientation. L/M
apparently drives the radio-quiet correlation only for
FWHM(\hb)$\simlt$4000 \kms\ which includes Narrow Line Seyfert 1
(NLSy1) galaxies and can be said to define an AGN ``main
sequence.'' Source orientation plays an increasingly important
role as FWHM(\hbbc) increases. We also argue that AGN lying
outside the radio-quiet ``main sequence,'' and specifically those
with optical Fe {\sc ii} much stronger than expected for a given
FWHM(\hbbc), may all be Broad Absorption Line QSOs.

\end{abstract}


\section{Introduction}

We have recently identified a correlation space for broad line AGN
involving:(1)  Balmer line width (FWHM(\hbbc)), (2) relative
strength of optical \feiiopt\  and \hb\ emission  lines (the
equivalent width ratio of \feii\ emission in the range 4435--4685
\AA\ and the broad component of \hb: W(\feiiq)/W(\hbbc)=\rfe), (3)
soft X-ray photon index \gs\ and (4) \civ\ broad line profile
centroid shift \citep{s2000a, s2000b}. In simplest terms the
parameters can be said to measure: (1) the broad line region
(BLR) velocity dispersion, (2) the relative strengths of low
ionization lines that are thought to arise in the same structure,
(3) the strength of a (thermal) soft X-ray photon excess, and (4)
the amplitude of systematic radial motions in the high ionization
gas. We call this parameter space ``Eigenvector 1'' (hereafter
E1) reflecting its partial origin in a principal component
analysis of the low redshift (z$\simlt$0.5) part of the
Palomar-Green quasar sample (Boroson \& Green 1992, hereafter
BG92). E1 allows us to discriminate between most AGN classes that
show broad emission lines \citep{s2000a,s2000b}.

E1 separates the majority of radio-quiet (RQ) sources from
radio-loud (RL) AGN. The E1 parameter space distribution also
suggests the possible existence of two RQ classes. Population A with
FWHM(\hbbc) $\simlt$ 4000 \kms\ and average E1 parameter values: (a)
\rfe $\approx$ 0.7, (b) \gs\ $\approx$ 2.8 and (c) \civ\ centroid
(blue)shift $\sim -800$ \kms. Population B includes all of the
remaining RQ AGN with FWHM(\hbbc) $\simgt$ 4000 \kms and shows
average E1 parameter values: (a) \rfe\ $\approx 0.4$, (b) \gs\
$\approx 2.3$\  and, (c) \civ\ centroid shift $\sim$ 0 \kms. See
\citet{s2000b} for sample variance and other details. RL and RQ
population B sources occupy a similar parameter domain in E1 and
show a large number of other observational similarities
\citep{s2000c}.

After reviewing the occupancy of the optical E1 parameter plane
(FWHM(\hbbc) vs. \rfe) and defining a ``main sequence'' for RQ
Population A, we show that the mean ionization level of the broad
lines decrease as one goes from RQ population  B to population A
sources (\S \ref{ion}). In \S \ref{iondec}, we show how   this
result  and the occupancy of the parameter plane can be explained in
terms of different values of the Eddington ratio (analyzed in \S
\ref{edd}) convolved with the effect of source orientation
(discussed in \S \ref{orient} and \ref{edd}). We also consider
potentially important outlier sources in \S \ref{out} and in
\ref{balqso}.

\section{Relevant Trends \label{trends}}

\subsection{Definition of ``Main Sequence'' and ``Outliers'' \label{out}}

Population A sources show a clear and significant correlation
among the E1 parameters while RQ population B and RL sources show
a larger scatter with no obvious correlation. RQ population B
sources occupy the same E1 parameter domain as, especially, flat
spectrum (core dominated) RL sources (see Table 2 in
\citet{s2000b}). It is however important to recognize that
uncertainties are larger for most measures of RL and RQ
population B sources. Limited S/N and line blending limit the
accuracy of \feii\ equivalent width measurements for sources with
W(\feiiq) $\simlt$ 20 \AA\ (\rfe $\simlt$0.2). That is why we
cannot rule out the possibility that the RQ population A
correlation extends into the RQ population B domain. At any rate,
the RL sources are preferentially found  in the same E1 domain as
RQ population B, and are rarely found in the domain of population
A. The concepts of a RQ population A -- population  B difference
and of a RQ population B -- RL similarity were motivated by the
optical parameters but they are reinforced by X-ray \gs\ and UV
\civ\ line shift measures. These differences/similarities are true
irrespective of the reality of a parameter space break between
population A and B  \citep{s2000b}.

\begin{figure}
\plotone{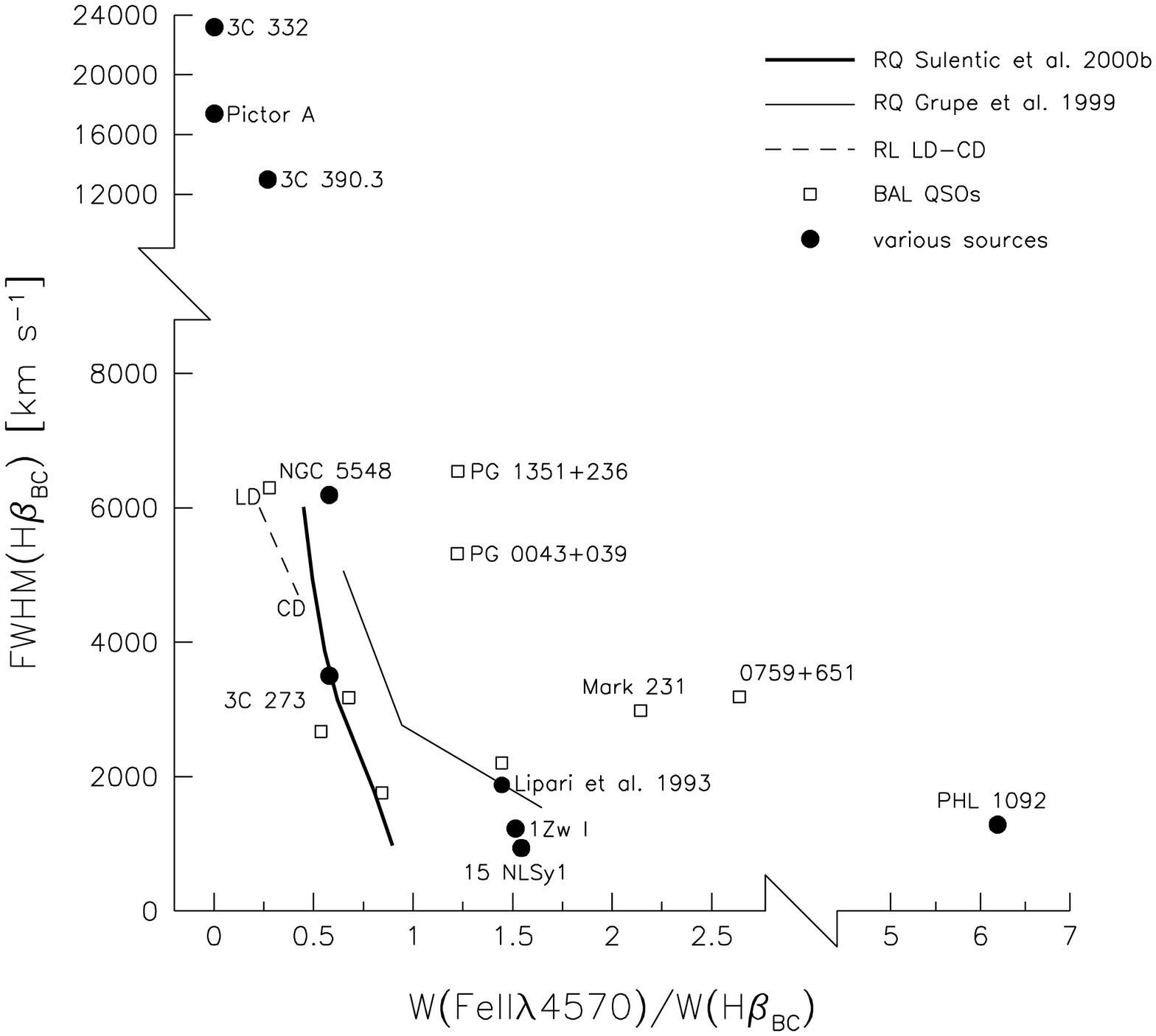} \figcaption[]{The  parameter plane FWHM(\hbbc) vs.
\rfe. The two solid lines mark the average positions of the sample
of \citet{grupe} and \citet{s2000b}. The dashed line traces the
average loci of core-dominated (CD)  and lobe-dominated (LD) RL AGN.
Open squares identify BAL QSOs. The five not labeled objects are:
PKS 1004+130, PG 1411+442, PG 2112+059, PG1011+054,  PG1700+518
 (in order of increasing \rfe).  Other relevant sources
(see text) are also plotted. The broken scale allows for the
inclusion of three wide-separation double-peaked RL AGN, and of PHL
1092. \label{fig01}}
\end{figure}

Figure \ref{fig01} presents a schematic view  of source occupancy in
the FWHM(\hbbc) vs. \rfe\ parameter plane.  We show correlation
trends for: (1) our sample \citep{s2000b}, (2) two radio-loud
samples \citep{b96,c97}, and  (3) a soft X-ray selected sample
\citep{grupe}. The solid lines representing (1) and  (3) connect
average values for sources in FWHM(\hbbc) ranges 0--2000, 2000--4000
and $\simgt$ 4000  \kms\ respectively. All samples show the same
general trends with broad line, \feii -weak RQ population B/RL
sources displaced towards the upper left and narrow line, \feii\
strong, population A RQ towards the lower right. These lines
indicate a tendency for AGN to lie along a ``main sequence'' (MS).
In addition to some prototype sources, we show data points for
\feii\  strong (FIR bright) quasars \citep{lipari1993}.  The
\citet{lipari1993} sources are shown as a single point representing
their average FWHM(\hbbc) and \rfe,  with the exception of two
objects, IRAS 0759+651 and Mark 231 which are reported individually
in Fig. \ref{fig01}. We also identify  all known broad absorption
line (BAL) QSOs  with absorption W(\civ) $\simlt -9$ \AA\  from the
\citet{s2000b} and \citet{lipari1993} samples. For PG quasars, data
on absorption W(\civ) were obtained from \citet{brandt2000}  except
for PG 1351+234, for which we measured the \civ\ absorption on the
IUE spectrum SWP54205.

We see that I Zw 1 (and other NLSy1), as well as several (low z) BAL
QSOs, are located toward the high \rfe\ end of the MS. One could
infer from Figure \ref{fig01} that the MS may extend from the
broadest double-peaked RL sources to the narrowest and \feii\
strongest sources (like PHL 1092). It is not yet clear if these
extrema should be considered outliers or extensions  of the
correlation found for the bulk of RQ Population A sources studied so
far (see  also \S \ref{disc}). We can identify one clear class of
outliers: sources occupying the upper right quadrant of the E1 plane
(Fig. \ref{fig01}, which turn out to be all BAL QSOs. These are
sources with FWHM(\hbbc) $\simgt$ 3000 \kms\ and \rfe $\simgt$ 1).
In general, sources with \rfe$ >>$ [FWHM(\hbbc)/(2500
\kms)]$^{-1.4}$\ (which approximately defines the upper boundary to
the MS of \citet{s2000a}) are sources with \feii\ in excess of the
mean value expected for a given FWHM(\hbbc). They will be briefly
considered in \S \ref{balqso}.

\subsection{An Ionization Decrease from population B to population A
\label{ion}}

Table 1 shows relevant mean parameter values for  RQ population A
and B sources as well as for the extreme population A NLSy1 sources.
All averages and sample standard deviations are from \citet{s2000b}
except for I(\siiii)/I(\ciii) which comes from HST data
\citep{laor94,laor95,wills1999}. The decrease in equivalent width of
\civ\ (our representative high ionization line) along with an
increase in W(\feiiq) (low ionization emission) suggest a systematic
decrease in ionization level from RQ population B to population A.
We interpret the data in Table 1 by considering the behavior of
I(\siiii)/I(\ciii), W(\civ) and W(\hbbc) as a function of the
ionization parameter (U) and electron  density \ne. We compare the
observed values with a grid of {\sc cloudy} computations for AGN
broad line emission \citep{kor97}. The models assumed a total column
density  $N_C \sim 10^{23}$ $cm^{-2}$ and a standard AGN continuum
(model 3--23 of \citet{kor97}). Figure 2 shows that all trends
passing from population B towards population A {\em consistently
suggest a decrease in U and an increase in $n_e$ (from log(U) $\sim$
$-$1 to $-$1.5, log($n_e$) $\sim$ 9.5, canonical value from AGN
photoionization models to log(U) $\sim$ $-$2 to $-$2.5, log($n_e$)
$\sim$ 11.5}). Notably, the density sensitive line ratio
I(\siiii)/I(\ciii) (almost independent of U for log(U) $\simlt$
$-$0.5) indicates that $\log$(\ne) $\sim$ 10.5 -- 11 towards the
NLSy1 domain.

These considerations quantify a general trend which is {\em very
appreciable} in the spectra of AGN with different Balmer line
widths (see Fig. 2 of \citet{s2000a}) and which has been
systematically ignored in photoionization computations. It is
probably the origin of our inability to explain some line ratios
in quasar spectra \citep{s2000b} and is obviously a zeroth-order
result. Our earlier comparison of \civbc\ and \hbbc\ properties
motivated us to suggest that high  and low ionization lines are
not emitted in the same region, at  least in population A sources
\citep{m96,d2000}.

\section{The Main Physical Parameters}

\subsection{The Role of Orientation in E1 \label{orient}}

Why is orientation important and why, at the same time, can it not
account for all of the phenomenology?  Important evidence in favor
of orientation effects in RQ AGN involves  \hbbc\  -- \civbc\
profile comparisons  \citep{m96, s2000b}. Some NLSy1 sources such as
I Zw 1 show a \civbc\ profile that is almost completely blueshifted
relative to a very narrow \hbbc. This robust observational result is
easily explained in terms of: (1)  a high-ionization wind emitting
\civbc\ and (2) an optically thick disk emitting \hbbc. The disk
will obscure the opposite (receding) side of the high-ionization
outflow \citep{m96}. Not all NLSy1 show large amplitude \civ\
blueshifts but the currently observed range is from 0 to $-$5000
\kms.  NLSy1 also show low W(\civ) \citep{rod}. The rarity of large
amplitude \civ\ blueshifts points toward a role for orientation
since \civ\ shifts are expected to be strongly orientation dependent
in a disk + wind scenario.

In principle, it is possible  to ascribe low W(\hbbc), W(\civ)
and W(\oiii) to an orientation dependent ``blue bump''  that
amplifies the UV/optical continua and whose contribution to the
continua increases with decreasing inclination \citep{m96}.
However, this assumption is rather {\it ad hoc} and can not
easily account for the increase of \rfe\ in sources with narrower
FWHM(\hbbc). In that scenario \rfe\ should  remain constant
because \hbbc\ and \feii\ would be similarly affected  by the
amplified continuum unless \hbbc\ and \feii\ are both, but
differently, anisotropic \citep{m96}. We remark that the observed
\rfe\ increase  appears to be mainly due to a decrease in
W(\hbbc) towards the NLSy1  domain. This has been interpreted as
the effect of collisional suppression of  \hb\ \citep{gas85}
implying an increase in electron density. The correlation
between  \ne-sensitive I(\siiii)/I(\ciii)\ ratio and FWHM(\hbbc)
also cannot be explained in  terms of orientation. Physical
conditions must change significantly along the AGN main sequence.

Further difficulties for an orientation-only hypothesis may
involve the  range of \oiiiopt\ luminosity observed among PG
quasars \citep{bg92}. We suggest that this issue needs
reconsideration also because the \oiiiopt\ emitting region in
several Seyfert galaxies shows a bipolar structure
\citep{falcke98}, a result that suggests a strong orientation
dependence.

\subsection{The Eddington Ratio: A Parameter Affecting BLR Physical
Conditions \label{edd}}

E1 shows us that the diversity of AGN properties can be organized
on the basis of a set of parameters involving: \gs, FWHM(\hbbc),
\rfe\ and \civ\ line shift. Apart from the reasons outlined
above, it is reasonable to infer that orientation alone would be
unlikely to account for the E1 correlations because we would
expect to find an overlapping domain for {\em all} RQ and RL AGN.
A distinct RL sequence, if it exists, is apparently displaced
from the RQ one. E1 suggests that it begins near
FWHM(\hbbc)$\sim$4000 \kms\ \citep{s2000a, s2000b}. Our BG92
dominated sample suggests that RQ sources become rare above
FWHM(\hbbc) $\sim$6000 \kms\ while RL are common from 4000 to, at
least, 8000 \kms. Support for a RQ-RL displacement comes from the
detection of a strong \ha\ line in BLLAC with FWHM(\ha)$\sim$4000
\kms\ \citep{corbett2000} which should be a near pole-on RL
source. The presence of the soft X-ray excess as one of the
principal correlates in E1 suggests that the Eddington ratio
(i.e., the ratio L/M) may be the most important physical
parameter driving E1, as well as the main factor accounting for
the RQ and RL sequence displacement
\citep{bg92,pounds95,boller96,s2000a}.

\subsection{A Correlation Between FWHM(\hbbc) and L/M? \label{correla}}

We often refer to the L/M ratio rather than the Eddington ratio
(ratio between the bolometric and  Eddington luminosities =
dimensionless accretion rate $\rm \dot{m} \propto L/M$) because we
rely, in this context, on independent observational measurements for
both L and M. X-ray variability determination of M for a few AGN
\citep{czerny} suggests an anti-correlation between L/M and
FWHM(\hbbc). Figure \ref{fig03} shows the best current observational
evidence for the anti-correlation between dimensionless accretion
rate and  FWHM(\hbbc). It involves the sources with most accurate
reverberation and/or X-ray variability based black hole mass
determination. Sources like NGC 4051 and 4151 show a transient broad
line component, which at some epochs is completely absent in
unpolarized light (see e.g., \cite{ulrich1997} and references
therein). FWHM(\hbbc) and X-ray properties for such sources may not
allow a reliable mass estimate that is comparable with AGN showing a
more typical variability pattern. At the other extreme, the two
highest $\dot{m}$\ radiators, PHL 1092, and IRAS 13224$-$3809 show
extreme optical properties (see \S \ref{disc} for a possible
interpretation). The \citet{czerny} dataset is obviously biased
towards objects that show large amplitude X-ray variability. Only
six sources remain in the X-ray sample if we omit them which is too
few for a reliable determination  of the correlation coefficient.
The results shown in Figure \ref{fig03} are suggestive with the
X-ray based points showing a correlation at a 2$\sigma$ confidence
level. If only the six sources are considered, Pearson's correlation
coefficient r$_P$\ is $\approx -0.85$ with a probability P$\approx$
0.04 that uncorrelated points could give rise to the computed r$_P$.
If all data points are taken into account, r$_P \approx -0.75$
(P$\approx$ 0.02).

Other approaches favor a strong correlation between $\dot{m}$\ and
FWHM(\hbbc). Modeling of a radiation pressure-driven wind
\citep{nicastro,witt} predicts the relationship between $\dot{m}$
and FWHM(\hbbc) (shown as the thick line of Fig. \ref{fig03}). A
rather strong correlation between L/M and FWHM(\hbbc) also emerges
using masses derived from reverberation mapping (\citet{kaspi2000})
(filled circles and squares in Fig. \ref{fig03}). If all RQ and RL
objects are considered except the outliers NGC 3227 (another
transient \hbbc\ object) and NGC 4051, the correlation coefficient
is r$_P \approx$ -0.75 (P$\approx$ 2$\times$10$^{-4}$). We obtain
the following functional relationship by a robust fitting technique
(e.g., \citet{nr}):

\begin{equation} \label{linea} (\frac{\lambda
L_\lambda}{M})_\odot \approx 6.2 \times 10^{3}
FWHM_{1000}(H\beta_{BC})^{-2},
\end{equation}

where L$_\lambda$\ is the specific luminosity at $\approx$ 5000
\AA\ and FWHM(\hbbc) is expressed in units of 1000 \kms. In order
to transform  to $\dot{m}$ we assume a constant bolometric
correction $\approx -2.5$, appropriate for the typical AGN
continuum as parameterized by \citet{mf87}. The
$\dot{m}$--FWHM(\hbbc) best fit is shown in Fig, \ref{fig03} as a
thin solid line.

A FWHM(\hbbc) -- L/M correlation in the case of reverberation masses
is not surprising since it results in part from circular arguments
(i.e., FWHM(\hbbc) is used to compute M from reverberation mapping
data). Monte Carlo simulations were carried out assuming: (1) that
distance r, FWHM(\hbbc) and luminosity are randomly distributed and
uncorrelated  in the observed ranges and (2) that M is related to
the velocity dispersion by the virial relationship: M $\propto$ r
v$^2$. Observational errors for $\log \dot{m}$ were assumed to
contribute a Gaussian scatter with $\sigma \approx 0.15$. In
approximately 4000 random trials we found a probability P$\simlt$
0.05 that a correlation coefficient as large as 0.75 would be
obtained. In order to fully circumvent the circularity issue we also
assumed that  FHWM(\hbbc) does not correlate strongly with either L
or M separately (as noted also by \citet{s2000a}). Optical
luminosity appears to be an orthogonal variable with respect  to the
E1 parameters (in BG92 it is part of  their Eigenvector 2). Mass and
FWHM(\hbbc) have r$_P \approx$0.48. If we  simulate data points for
which: (1) the correlation between FWHM(\hbbc)  and M falls in the
range  $0.43 \simlt r_P \simlt 0.53$ and (2) mass and luminosity are
not correlated, we obtain a negligible probability that an $r_P
\approx $0.75 correlation  between FWHM(\hbbc) and $\log \dot{m}$
could occur randomly.

We conclude that the circularity inherent in the mass computation
cannot fully explain the strength of the observed L/M vs.
FWHM(\hbbc) correlation for reverberation masses. This provides
much needed support for a physical relationship between L/M and
FWHM(\hbbc). Given the caveats  outlined above (small numbers of
X-ray determined masses;  M(FWHM) dependence in optical
reverberation masses), an independent verification of any
$\dot{m}$ -- FWHM(\hbbc) relationship would be best derived from
X-ray variability (Braito \& Marziani, in preparation). In the
present study, we  {\em assume} that Eq. \ref{linea} describes the
appropriate relation.

Differences between the disk + wind model expectation and the
linear fit described by  Eq. \ref{linea} are appreciable  for
FWHM(\hbbc) $\simlt$ 1500 \kms\ (Fig. \ref{fig03}).  The disk +
wind model predicts an average difference of more than one order
of magnitude between NLSy1 and other population A sources. The
assumption of constant $\dot{m}$\ for all population A yields a
lower normalized $\chi^2_\nu$\ ($\approx$ 3.2) than the disk +
wind fit ($\chi^2_\nu \approx 4.3$) (see also \S \ref{disc} for
possible interpretations).

\section{Why Does Ionization Level Decrease With Increasing L/M? \label{iondec}}

The ionization parameter can be defined as
\begin{equation}
U = \frac{Q(H)}{4 \pi r^2 n_e c},
\end{equation}

where Q(H) is the number of hydrogen ionizing photons and r is the
distance of the BLR from the central continuum source. U can be
rewritten in terms of L/M and M, if we assume:
\begin{enumerate}

\item $Q(H)\approx  f L_{bol} /<h\nu>  $. A typical AGN continuum
as parameterized by \citet{mf87} yields $<\nu>\approx 9.96 \times
10^{15}$Hz and f $\approx$ 0.54.
\item The velocity field for the low
ionization line-emitting gas is  mainly rotational. We assume
that the velocity dispersion is the square root of the mean
square velocity for a rotating annulus:
\begin{equation}
\sigma = < v^2 >^{1/2} = \frac{1}{2 \sqrt{2}} \sqrt{\frac{GM}{r}},
\end{equation}
with FWHM(\hbbc) = 2.35 $\sigma$.
\end{enumerate}
We also consider that the ratio I(\siiii)/I(\ciii) is a good
density diagnostic (almost independent of the ionization
parameter) in the density range $ 9.5 \simlt \log(n_e) \simlt
12$.  {\sc cloudy} \citep{cloudy} photoionization computations
suggest that
\begin{equation}
\frac{ I({\rm Si~  III]}\lambda 1892)}{ I({\rm  C ~ III]} \lambda
1909)} \approx -3.91 + 0.41 \log n_e. \label{density}
\end{equation}
Since the ratio I(\siiii)/I(\ciii) is  directly correlated with
FWHM(\hbbc) \citep{wills1999}, adopting the  FWHM(\hbbc) -- L/M
correlation yields:
\begin{equation} \log n_e \approx 11.1 - 1.33 \log
FWHM_{1000}(H(\beta) \approx 7.72 + \frac{2}{3} \log ({\rm L_{\rm
bol} / M})_\odot. \label{nelm}
\end{equation}

We can use Eq. \ref{nelm} to write U in terms of L$_{bol}$/M:

\begin{equation}
U = 0.26 (\frac{L_{bol}}{M})_{\odot,4}^{-(1+x)} M_{\odot,7}^{-1},
\end{equation}

with $x = 0.67$, mass in units of 10$^7$$M_\odot$, and the
luminosity-to mass ratio in units of 10$^4$\ the solar value
(L/M$_\odot \approx$ 1.9 ergs s$^{-1}$ g$^{-1}$). This accounts for
the somewhat counterintuitive result that U decreases with
decreasing FWHM(\hbbc) and increasing L/M. We note that for
$M_{\odot,7} = 1$ we get $\log U \approx -2$\ for
FWHM(\hbbc)$\approx$ 2000 \kms\ as expected. However a shallower
dependence may be possible; the L/M power is constrained within 1.1
$\simlt 1 + x \simlt$ 1.7 for a reasonable choice of the input
parameters. For instance, if we use the relationship between
luminosity and BLR radius $r_{BLR} \propto L^{0.7}$ \
\citep{kaspi2000}, we would obtain $x \approx 0.1$. Values of  $1 +
x \simgt 1 $\ are a consequence of the assumption of a Keplerian (or
virial or similar) velocity field (i.e., v $\propto 1/\sqrt{r}$).

\section{Connecting the Observational Plane to L/M and i
\label{conn}}

The above results allow us to relate L/M to the FWHM(\hbbc) -- \rfe\
plane of E1. Determining the relationship between \rfe\ and U is not
trivial since \feii\ emission is poorly understood. Photoionization
calculations suggest a four-fold increase in total \feii\ emission
from $\log U \approx -1 $ to $-$2. We assume that \rfe\ scales as
total \feii\ intensity divided by I(\hbbc) as a function of U for an
average density \ne\ $\approx$ 10$^{10}$ \cm3. We use  the
calculations of \citet{kor97} to estimate the dependence on U. The
normalization has been chosen following \citet{netzer} with \rfe\
$\approx$ 0.25 for $\log U= -1$. We note that using the possibly
stronger correlation between I(\siiii)/I(\ciii) and \rfe\ (i.e.,
I(\siiii)/I(\ciii) $\approx$ 0.1 + 0.5 \rfe,  \citet{wills1999})
gives a consistent relationship without any assumption about the
relationship between U and \rfe. Equating the above relationship
involving I(\siiii)/I(\ciii) and \rfe\ to Eq. \ref{density} and using
Eq. \ref{nelm} to relate \ne\ and L/M allows us to obtain  \rfe\
$\propto 0.55 \log {L/M}$.

If low ionization lines like \hb\ are emitted in a flattened
configuration, then some  effect of viewing angle is expected. In
order to take into account the effect of orientation we assume
that:

\begin{enumerate}
\item
 the relationships employed above are valid for an average $<
i > \approx 30^\circ$. We write the FWHM(\hbbc) dependence on i as

\begin{equation}
\rm FWHM(i) = FWHM(0) + \Delta FWHM \cdot \sin i,
\end{equation}

where \begin{equation}\rm \Delta FWHM = 2 [ FWHM(i=30^\circ,
\frac{L}{M}) - FWHM(0) ].
\end{equation}

\item \rfe\ depends on i following a $\sec(i)$ law with a ratio
\rfe\ that may change by a  factor 1.6 (an amplitude taken from
the mean \feii\ difference between lobe-dominated and
core-dominated RL objects).
\end{enumerate}

These assumptions allow us to reproduce the parameter space covered
by sources in our samples by assuming $\log M \sim 8$\ in the
expression of U, and FWHM(0) = 500 \kms. Fig. \ref{fig04} \ shows a
grid of theoretical values superimposed on the data points of
\citet{s2000a}. If  $1 + x < 1.7$ ($1 + x = 1.7$ is assumed  in Fig.
\ref{fig04}), then $\dot{m}$\ somewhat larger will result for the
same \rfe, yielding however the same qualitative behavior. For
instance, for $x \approx$ 0.4, $\log (L/M)_\odot \approx$ 4.5
corresponds to \rfe $\approx$ 1.5.

\section{Discussion \label{disc}}

Our calculations do not attempt to reproduce the observed point
distribution, but only to account for the occupancy of the parameter
plane, since instrumental factors and biases affect the
distribution. The cluster of points at \rfe $\approx$0.2, for
example, is due to limits on S/N and resolution. Another source of
concern involves the role of selection biases in our E1 AGN sample.
RQ population A sources are favored by soft X-ray (e.g.,
\citet{grupe}) and optical color-based (e.g., BG92) selection
techniques while  RL/RQ population B AGN are not. The latter sources
may be seriously under-represented in the \rfe\ vs. FWHM(\hbbc)
plane.

On the theoretical side, a distribution of masses will  blur the
grid, since different masses would deform and displace the grid
horizontally. An additional source of  scatter may involve Fe
abundance. Therefore no rigorous inference can be made about
individual values of L/M and i from the Figure \ref{fig04} grid.

The low \rfe\ region (\rfe $\simlt 0.5$) of Figure \ref{fig04}
suggests that orientation is responsible for some of the population
B sources. They would fall in the population A domain if viewed
face-on. Towards the middle of Figure \ref{fig04}, we see that
decreasing i and increasing L/M apparently have a concomitant effect
(decreasing i  implies decreasing FWHM(\hbbc) and \rfe; increasing
L/M implies the same observational trends). This is also true  in
the domain of NLSy1, which should be an  L/M extremum (but not
necessarily of i). The concurrent effects of both parameters may
explain why the correlations above have been found by a number of
workers without any contradictory result. Observational prediction
of some wind models are sensitive to both i and L/M, and account
for, at least qualitatively, the \civ\ shift amplitude distribution
as considered by \citet{s2000b}.

Radiation pressure driven wind models \citep{nicastro,witt} predict
a decrease of $\dot{m}$\ with increasing FWHM(\hbbc). However, the
model by \citet{nicastro} predicts highly super-Eddington accretion
for NLSy1. Existing evidence is still sparse, but neither a
dynamical mass determination nor X-ray mass estimate supports this
prediction. Rather, NLSy1 as the bulk of population A sources, seem
constrained within $0.3 \simlt \dot{m} \simlt 1.0$ (e.g.,
\citet{puchna2000} constrain $0.3 \simlt \dot{m} \simlt 0.7$\ for RE
J1034+396; a similar result is inferred for Akn 564
\citep{comastri2000}. \citet{laor2000a} also infers $\dot{m} \simgt$
0.3  for NLSy1 from X-ray variability. RQ AGN with reverberation
mapping mass estimates (excluding NGC 4051 and 4151) of Fig.
\ref{fig03} show that there is no strong discontinuity between NLSy1
and the rest of population A (i.e., sources with 2000 \kms\ $<$
FWHM(\hbbc) $\le$ 4000 \kms). The $\dot{m}$\ values become
significantly different if population A and population B are
compared with a K-S test (the difference is also appreciable in Fig.
\ref{fig03}). The K-S test applied to population A (13 sources) and
population B (11 sources) yields a probability P$\approx$1$\times
10^{-3}$ that the $\dot{m}$ values are drawn from the same parent
population.

Most NLSy1 sources radiate at $\log \dot{m} \approx 0.0$, as do a
sizeable fraction of the sources with 2000 \kms\ $<$ FWHM(\hbbc)
$\le$ 4000 \kms\, supporting the identification of a unique
population up to at least FWHM(\hbbc)$\approx$ 3500 \kms. PHL
1092 may be radiating at $\dot{m} \simgt 10$\ (IRAS 13224$-$3809
may be another case). The extreme location of PHL 1092 in Fig.
\ref{fig01}\  and in Fig. \ref{fig03}\ is consistent with both an
exceptionally high value of $\dot{m}$ ($\gg 1$) and  a pole-on
orientation (also suggested by the strong X-ray variability
\citep{forster}). Sources like PHL 1092 may therefore be
intrinsically rare even if not peculiar in a strict sense.

Recent results suggest a clear dichotomy between RL and RQ AGN in
terms of black hole mass  with RL AGN having a systematically
larger black hole mass \citep{laor2000b}. According to our
considerations, the optical properties of AGN should be largely
transparent to black hole mass differences (with a dependence of
U on M yielding a second order effect). However, the probability
of having low L/M is obviously favored for large masses. This is
in agreement with the upwardly displaced location of RL AGN in
the \rfe\ -- FWHM(\hbbc) plot with respect to radio quiet AGN.
Also, in the idealized case of a sample where L $\approx$
constant, there could be a sequence of increasing mass  from
NLSy1 to population A and then to population B,  with
 RL AGN hosting the most massive black holes.

\subsection{On the Nature of the Outliers \label{balqso}}

A major difference we are able to identify between MS and outlying
BAL QSOs  is related to their far IR spectral index $\alpha$\
($f_\nu \propto \nu^{-\alpha}$) between 25 and 60 $\mu$m. The index
$\alpha$(25,60) is 1.47 and 1.32 for Mrk 231, and 0759+651
respectively. Consistently, from ISO data \citep{haas2000},
$\alpha$(25,60)$\approx$ 1.13 for PG 0043+039 (PG 1351+236 has been
detected only at 60$\mu$m). In both Fig. \ref{fig01}\ and in Fig.
\ref{fig04}\ all the BAL QSOs known to us from \citet{s2000a} and
\citet{lipari1993} with W(\civ)$\simlt -9$ \AA\ in absorption are
reported. The three MS BAL QSOs for which there are reliable IRAS or
ISO data (PG 1411+442, PG 1700+518, and PG 1001+054) have
$\alpha$(25,60)$\approx -$0.24, 0.56, $-$0.33 respectively. In the
remaining  two cases (PG 1004+130 and PG 2112+059) it is not
possible to compute the $\alpha$(25,60), but detection at 12$\mu$m
and no detection at 60 $\mu$m argues against large $\alpha$(25,60).

The most straightforward interpretation of this difference is a
continuum steeply rising toward the far IR due to a significant
contribution from circum-nuclear star formation in the outlying
BAL QSOs (in line with the analysis of \citet{haas2000} of the
far IR spectral shapes of PG quasars). This contribution  may not
be dominant in the MS BAL QSOs. For both  0759+651 and Mrk 231,
several lines of evidence suggest the presence of a strong
circumnuclear starburst affecting the integrated broad line
spectrum of these AGN \citep{taylor1999,lipari1994}. If we
consider the general population of AGN in a diagram rest-frame
W(\feiiq) vs. W(\hbbc), we see that \feii\ strong quasars (their
average is shown in Fig. \ref{fig01}, \citet{lipari1993}) define
a boundary with W(\feiiq)$\approx$ W(\hbbc), while the general
population of AGN fills the area with W(\feiiq) $<$ W(\hbbc). Mrk
231 and 0759+651 remain outliers: they show W(\feiiq)
significantly larger than that expected from W(\hbbc). This may
indicate a significant overproduction of \feii\ due to an
additional excitation mechanism (i.e., shocks), possibly
associated with strong circumnuclear star formation.

\section{Conclusion}

Our work (\citet{s2000a} and references therein) shows that
optical \feii\ emission  is a fundamental parameter in AGN
correlation studies. This result underlies the need for more
sophisticated models for the production of  \feii\ (e.g.,
\citet{verner}). It also points out the need for much higher S/N
spectroscopic observations of many sources with weaker \feii\
emission in order to clarify the upper left part of E1. In this
paper we have attempted to explain, semi-quantitatively but self-
consistently, the diversity among RQ AGN. Our attempt has been
based upon the assumption that two of their most prominent
emission features, \hbbc\  and \feii, are influenced primarily by
source orientation and the Eddington ratio. The results have
allowed us to reinforce the distinction between two RQ AGN
populations and to tentatively identify those AGN that may be
peculiar in a statistical and phenomenological sense including
some BAL QSOs.

\acknowledgements

The authors acknowledge  fruitful discussion and encouragement
from M. - H. Ulrich. MC, PM and JS acknowledge support from the
Italian Ministry of University and Scientific and Technological
Research (MURST) through grants Cofin 98$-$02$-$32 and Cofin
00$-$02$-$004. TZ acknowledges support from the Slovene Ministry
of Research and Technology.

\clearpage

\begin{figure}\epsscale{0.45}
\plotone{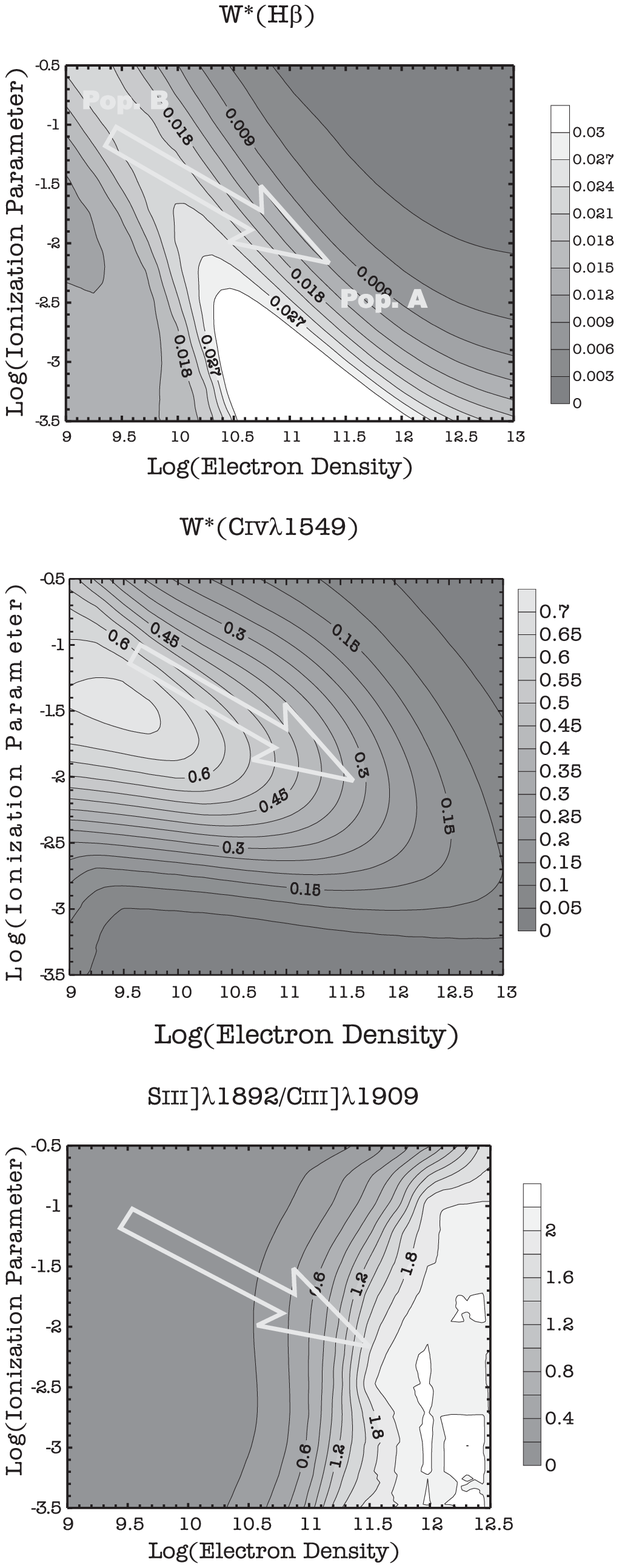}
\figcaption[e1_fig02.eps]{Behavior of the ratio
I(\siiii)/I(\ciii), of normalized W(\civ), and of normalized
W(\hb) as a function of electron density \ne\ and ionization
parameter computed with {\sc cloudy} by \citet{kor97}. Arrow
tails are placed roughly at the value of (\ne, U) expected for
Population B sources, arrow head at that for Population A.
Equivalent width normalization is by continuum at 912 \AA\ (see
\citet{kor97} for further details). \label{fig02}}
\end{figure}

\begin{figure}\epsscale{1.0}
\plotone{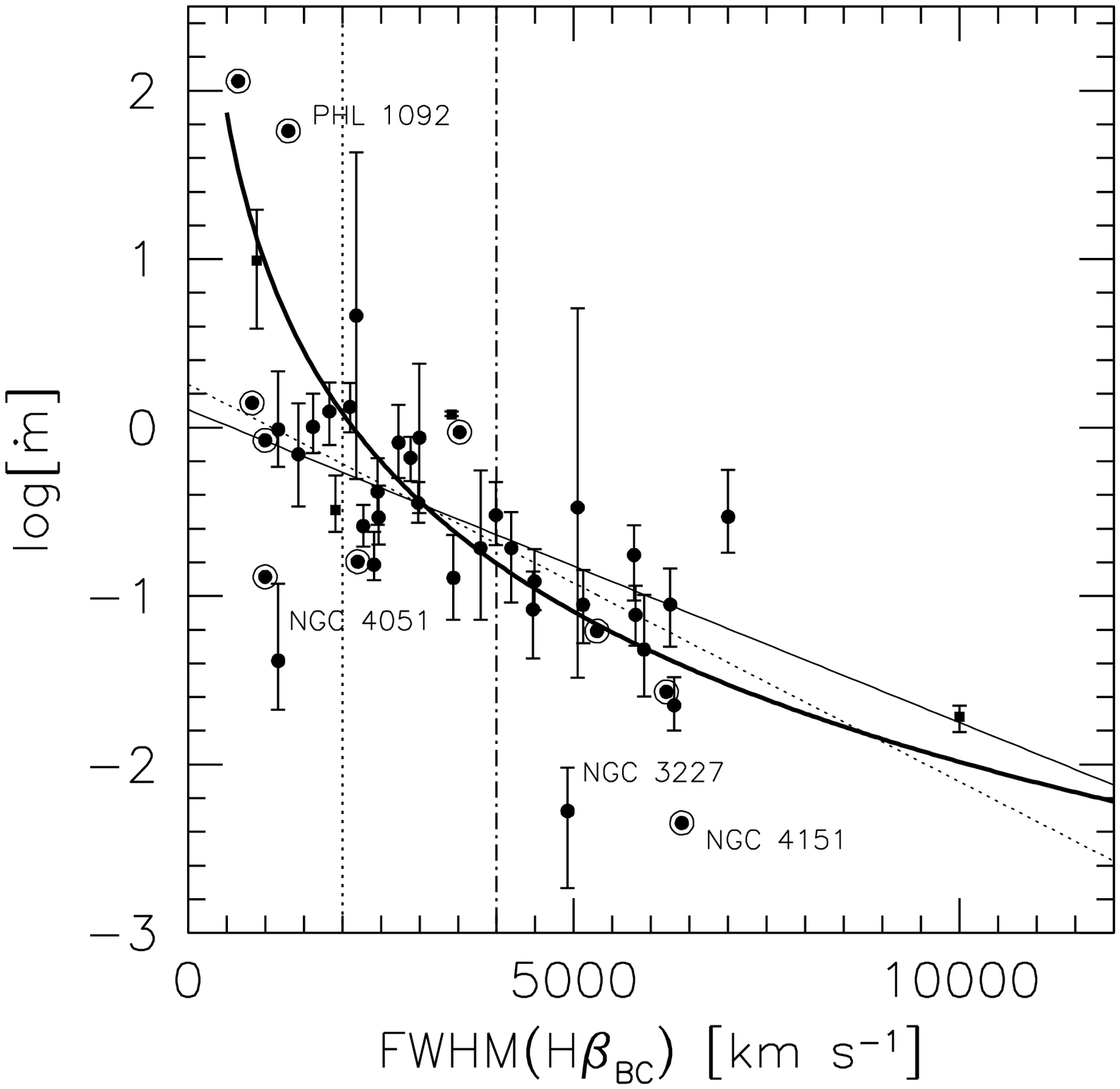}
 \figcaption[e1_fig03.eps]{Relationship
between dimensionless accretion rate  and FWHM(\hbbc). Filled
circles and squares are RQ and RL AGN, respectively, with
reverberation mapping mass estimates  from \cite{kaspi2000}. Ringed
filled circles label AGN with X-ray variability mass determination
\citep{czerny}. The thick line marks the prediction of a disk + wind
model \citep{nicastro}. The thin line is the best fit employing a
robust techniques for all data points of \citet{kaspi2000}; the thin
dotted line represents the same with the exclusion of radio loud
AGN. The vertical lines mark the boundaries of NLSy1 (dotted) and
population A nuclei (dot-dashed). \label{fig03}}
\end{figure}

\begin{figure}
\plotone{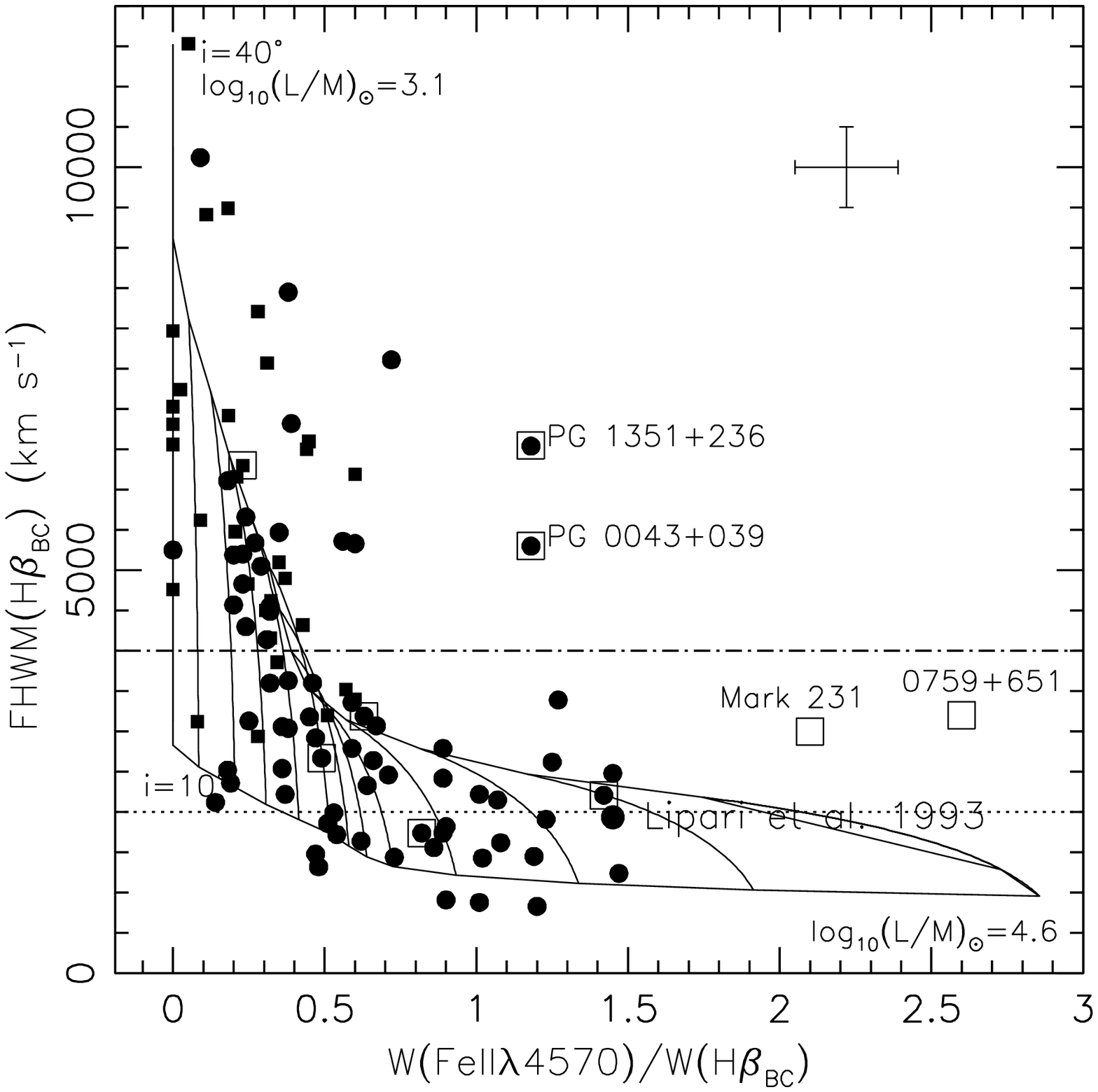} \figcaption[]{The FWHM(\hbbc) vs. \rfe\ with
superimposed a grid of theoretical values as a function of i ($
10^\circ \le i \le 40^\circ$) and L/M, expressed in solar values,
for $ 3.1 \le \log \frac{L}{M}_\odot \le 4.5 (\Rightarrow \log
\dot{m} \approx 0)$, at steps of $\log \frac{L}{M} = 0.1$.  A value
of $\log M \sim 8 $ in solar units has been assumed for U. Data
points are from  \citet{s2000b}; as for Fig. \ref{fig03}, filled
circles represent RQ, and filled squares RL AGN. Data from
\citet{lipari1993} are shown as an average data point for objects
clustering at \rfe $\approx$ 1.5 and FWHM(\hbbc) $\sim$ 2000. Open
squares mark BAL QSOs; they are traced around a filled circle if the
object was present in the sample of \citet{s2000b}. The horizontal
lines mark the boundaries of NLSy1 (dotted) and population A nuclei
(dot-dashed). The error bars in the upper right corners of the
Figure are typical 2$\sigma$\ confidence level errors for a
data-point at FWHM(\hbbc) $\approx$ 4000 \kms, and \rfe $\approx$
0.5. \label{fig04}}
\end{figure}

\begin{deluxetable}{lcccccc}
\tablewidth{0pt} \tablecaption{Emission Line Mean Parameter Values
for Radio Quiet AGN Populations }

\tablehead{& \colhead{W(\hbbc)} & \colhead{W(\feiiq)} &
\colhead{\rfe} & \colhead{I(\siiii)/I(\ciii)}  &
\colhead{W(\civ)} \\
& \colhead{[\AA]} & \colhead{[\AA]} &  & & \colhead{[\AA]}  }
\startdata

NLSy1  &  72$\pm$28 &54$\pm$17  &0.8$\pm$0.3&
0.53$\pm$0.16&  41$\pm$19\\
%Pop A - NLSy1 &  2-4000 & 110$\pm$38 & 64$\pm$27  &
%&0.7$\pm$0.4 & 0.39$\pm$0.20\\
Pop A  &  95$\pm$39&   60$\pm$16  & 0.5$\pm$0.3&
0.43$\pm$0.19&   50$\pm$25\\
Pop B &  104$\pm$34 & 31$\pm$17  & 0.3$\pm$0.1 & 0.23$\pm$0.11 &  107$\pm$85\\
\enddata
\end{deluxetable}

\end{document}